\documentclass[nofootinbib,onecolumn,amsmath,notitlepage,preprintnumbers]{revtex4-1}
\usepackage{multirow}

\usepackage{setspace,nicefrac} 
\usepackage{amssymb,esvect,amsmath,graphicx,latexsym,amsthm,slashed,eso-pic,mathtools}

\newcommand{\ie}{{\it i.e.}}

\newcommand{\beq}{\begin{equation}} \newcommand{\eeq}{\end{equation}}
\newcommand{\bea}{\begin{eqnarray}} \newcommand{\eea}{\end{eqnarray}}

\newcommand{\lsim}{\mathrel{\hbox{\rlap{\lower.55ex\hbox{$\sim$}} \kern-.3em \raise.4ex \hbox{$<$}}}}
\newcommand{\gsim}{\mathrel{\hbox{\rlap{\lower.55ex\hbox{$\sim$}} \kern-.3em \raise.4ex \hbox{$>$}}}}

\newcommand{\mpl}{M_\mathrm{Pl}}

\def\lsim{\mathrel{\raise.3ex\hbox{$<$\kern-.75em\lower1ex\hbox{$\sim$}}}}
\def\gsim{\mathrel{\raise.3ex\hbox{$>$\kern-.75em\lower1ex\hbox{$\sim$}}}}

\newcommand{\Eq}[1]{Eq.~(\ref{#1})}

\newcommand{\be}{\begin{eqnarray}}
\newcommand{\ee}{\end{eqnarray}}

\newcommand{\benum}{\begin{enumerate}}
\newcommand{\eenum}{\end{enumerate}}
\newcommand{\bi}{\begin{itemize}}
\newcommand{\ei}{\end{itemize}}

\newcommand{\mbh}{  M_{\rm BH}  }

\begin{document}

\preprint{FERMILAB-PUB-20-471-T}

\title{GUT Baryogenesis With Primordial Black Holes}

\author{Dan Hooper$^{a,b,c}$}
\thanks{ORCID: http://orcid.org/0000-0001-8837-4127}

\author{Gordan Krnjaic$^{a,c}$}
\thanks{ORCID: http://orcid.org/0000-0001-7420-9577} 

\affiliation{$^a$Fermi National Accelerator Laboratory, Theoretical Astrophysics Group}
\affiliation{$^b$University of Chicago, Department of Astronomy and Astrophysics}
\affiliation{$^c$University of Chicago, Kavli Institute for Cosmological Physics}

\date{\today}

\begin{abstract}

In models of baryogenesis based on Grand Unified Theories (GUTs), the baryon asymmetry of the universe is generated through the CP and baryon number violating, out-of-equilibrium decays of very massive gauge or Higgs bosons in the very early universe. Recent constraints on the scale of inflation and the subsequent temperature of reheating, however, have put pressure on many such models. In this paper, we consider the role that primordial black holes may have played in the process of GUT baryogenesis. Through Hawking evaporation, black holes can efficiently generate GUT Higgs or gauge bosons, regardless of the masses of these particles or the temperature of the early universe. Furthermore, in significant regions of parameter space, the black holes evaporate after the electroweak phase transition, naturally evading the problem of sphaleron washout that is normally encountered in GUT models based on $SU(5)$. We identify a wide range of scenarios in which black holes could facilitate the generation of the baryon asymmetry through the production and decays of GUT bosons.

\end{abstract}

\maketitle

\section{Introduction}

One of the most significant open questions in cosmology is how the universe came to contain more matter than antimatter. This asymmetry is typically characterized in terms of the dimensionless quantity: 
\be
\eta \equiv \frac{ n_B-n_{\bar{B}}  }{n_{\gamma}} ,
\ee
where $n_B$ and $n_{\bar{B}}$ are the number densities of baryons and anti-baryons, respectively, and $n_{\gamma}= 2\zeta(3)T^3/\pi^2$ is the number density of photons in a blackbody distribution of temperature, $T$. It is also sometimes useful to write this quantity in terms of the number density of baryons per unit entropy density:
\be
Y_B \equiv \frac{n_B - n_{\bar B}}{s}  =   \frac{ 45 \zeta(3)  \eta  }{   \pi^4 g_{\star S}    } 
\approx 0.14 \eta,
\ee
where $g_{\star S} \approx 3.9$ is the number of degrees-of-freedom in entropy at the universe's current temperature. Although the value of this quantity could, in principle, evolve with time, measurements of the cosmic microwave background (CMB) and the primordial light element abundances each yield $Y_B \approx 8.8 \times 10^{-11}$ \cite{Zyla:2020zbs}, indicating that any primordial abundance of antimatter had already disappeared by the onset of Big Bang nucleosynthesis (BBN). This, in turn, indicates that some mechanism must have driven the early universe to transition from a matter-antimatter symmetric state (with $\eta=0$) at the end of inflation (or other initial condition), to one with an appreciable value of $\eta$.\footnote{In principle, it is possible for initial conditions to supply a baryon asymmetry that survives inflationary dilution, but this requires 
 exponential fine tuning and trans-planckian scalar field values~\cite{Krnjaic:2016ycc}.} While the nature of this mechanism remains unknown, it is well established that it must include each of the following elements, commonly known as the Sakarov conditions \cite{Sakharov:1967dj}:
\begin{enumerate}
\item{Interactions which violate the conservation of baryon number.}
\item{Interactions which violate the symmetries of charge (C), and charge-parity (CP).}
\item{Departure from thermal equilibrium.}
\end{enumerate}

Many proposals have been put forward which satisfy these conditions and could potentially explain the existence of the observed baryon asymmetry. Some of the most well-known examples include electroweak baryogenesis~\cite{Klinkhamer:1984di,Kuzmin:1985mm,Arnold:1987mh,Arnold:1987zg,Khlebnikov:1988sr,Kajantie:1996mn,Riotto:1999yt,Cline:2006ts}, GUT baryogenesis~\cite{Harvey:1981yk,Weinberg:1979bt,Nanopoulos:1979gx,Ignatiev:1978uf,Yoshimura:1979gy,Riotto:1998bt}, leptogenesis~\cite{Fukugita:1986hr,Buchmuller:2004nz,Buchmuller:2005eh,Pilaftsis:2005rv}, and the Affleck-Dine mechanism~\cite{Affleck:1984fy,Dine:2003ax}. Some of these ideas (namely leptogenesis and the Affleck-Dine mechanism) have not been significantly impacted by existing observations or experimental results, and remain as viable today as they were at the time of their proposal. In contrast, recent data has begun to place tension on many models of electroweak baryogenesis and GUT baryogenesis, in some cases even ruling out such scenarios.

Models of electroweak baryogenesis generally require the existence of relatively light particles in order to facilitate the necessary departure from thermal equilibrium. The lack of evidence for such particles at the Large Hadron Collider (LHC) significantly constrains this class of scenarios~\cite{Menon:2009mz,Cohen:2012zza,Curtin:2012aa,Carena:2012np,Krizka:2012ah,Delgado:2012eu,Huang:2014ifa,Kozaczuk:2014kva}, typically requiring fine-tuning and/or large couplings which lead to Landau poles at scales not far above $\sim$\,1 TeV~\cite{Cline:2017jvp,Chung:2012vg} (for potentially viable alternatives, see Refs.~\cite{Carena:2019une,Chiang:2019oms,Kozaczuk:2019pet,Angelescu:2018dkk,Chiang:2018gsn,Akula:2017yfr,Kobakhidze:2015xlz,Blinov:2015vma,Curtin:2014jma,Kurup:2017dzf}).
%
%
Meanwhile, theories of GUT baryogenesis have come under pressure from a very different direction, as a consequence of measurements of the CMB. In such scenarios, the baryon asymmetry is generated through the C, CP, and baryon number violating, out-of-equilibrium decays of very massive gauge or Higgs bosons in the very early universe. Although the precise mass of these exotic particles depends on the GUT gauge group and the representations under consideration, they are typically near the scale of GUT symmetry breaking. In supersymmetric $SU(5)$, for example, the GUT gauge bosons (denoted by $X$ and $Y$) have masses of approximately $M_{X,Y}\simeq 10^{16}$ GeV (in the absence of supersymmetry, GUTs based on $SU(5)$ are ruled out by constraints on the proton lifetime)~\cite{Amaldi:1991cn,Amaldi:1991cn,Langacker:1991an}. While the masses of the Higgs bosons found in GUTs are more sensitive to the choice of representation and other model-dependent considerations, the baryon number violating Higgs bosons in the simplest, 5-dimensional representation of $SU(5)$ are typically taken to be $M_H \sim10^{14}$ GeV or greater, an expectation that is additionally reinforced by constraints on proton decay.

In order for GUT baryogenesis to produce the observed baryon asymmetry, some combination of these very massive particles must have been copiously produced in the early universe. Constraints on the tensor-to-scalar ratio from Planck and the BICEP2/Keck Array, however, indicate that the energy scale of inflation was less than $V^{1/4}_{\rm Inf} < 1.6 \times 10^{16} \, {\rm GeV}$~\cite{Akrami:2018odb}. When this information is combined with the small magnitude of temperature fluctuations observed in the CMB, it suggests that 
%
the universe was reheated after inflation to a temperature of $T \lsim 10^9-10^{13} \, {\rm GeV}$~\cite{Akrami:2018odb,Bezrukov:2011gp,Bezrukov:2007ep}, well below the scale of grand unification, highly suppressing the    
thermal production of the baryon number and CP violating particles, and posing significant problems for many otherwise well-motivated theories of GUT baryogenesis.\footnote{Alternatively, massive particles could have been produced efficiently through a broad parametric resonance during the period immediately preceding conventional reheating (\ie~preheating)~\cite{Dolgov:1989us,Traschen:1990sw,Kofman:1994rk,Kofman:1997yn,Kolb:1996jt}.} 

Another challenge faced by many models of GUT baryogensis comes from the influence of sphalerons on any baryon asymmetry that might be produced. As soon as the temperature of the universe drops below $T\sim10^{12} \, {\rm GeV}$, sphaleron transitions rapidly drive the value of the net baryon-plus-lepton number ($B+L$) toward zero, erasing any asymmetry that may have been present at high temperatures~\cite{Riotto:1998bt}. The traditional way of avoiding this problem in GUT baryogenesis is for the particle decays to produce a non-zero value of $B-L$, which is not eroded by sphalerons. In many GUT models, however, such as those based on $SU(5)$, no such net $B-L$ is generated, leading to a situation in which the baryon asymmetry is efficiently washed out well before the electroweak phase transition.

It is within the context of these challenges for GUT baryogenesis that we consider the role that primordial black holes may have played in this process. This possibility has been discussed previously by a number of authors, in particular by Barrow {\it et al.}~\cite{Barrow:1990he}, and more recently in Refs.~\cite{Baumann:2007yr,Morrison:2018xla} (for other related work, see Refs.~\cite{Hawking:1974rv,Carr:1976zz,Zeldovich:1976vw,Toussaint:1978br,Turner:1979bt,Harvey:1990qw,Grillo:1980rt,Khlopov:1980mg,Hawking:1982ga,Polnarev:1986bi,Majumdar:1995yr,Upadhyay:1999vk,Bugaev:2001xr,Hook:2014mla,Hamada:2016jnq,Fujita:2014hha}). In this study, we focus on the ways in which a population of primordial black holes could serve to relax some of the constraints and challenges that are otherwise faced in many baryogenesis scenarios based on GUTs. In particular, the presence of black holes in the early universe could have impacted the process of GUT baryogenesis in at least three separate and important ways:
\begin{enumerate}
\item{ {\bf Heavy Particle Production:}
As black holes undergo Hawking evaporation, their temperature increases. Thus even the heaviest of particles can be produced by black holes, regardless of the temperature of the universe after inflation.
}
\item{ 
{\bf Avoiding Washout:}
For any black holes heavier than $\sim (3-6) \times 10^5 \, {\rm g}$ (but lighter than $\sim 5 \times 10^8 \, {\rm g}$ in order to evade constraints from BBN~\cite{Keith:2020jww,Carr:2009jm}), the process of Hawking evaporation would be completed only after the electroweak phase transition, thereby automatically avoiding any problems with sphaleron washout.}
\item{
{\bf Out of Equilibrium Decays: }
Any baryon number violating particles that are radiated from a black hole will be out-of-equilibrium with the thermal bath. Thus, their baryon number violating decays automatically satisfying the third Sakarkov condition, even if they decay promptly.
}
\end{enumerate}

The remainder of this paper is structured as follows. In Sec.~\ref{sec:bh}, we review the formation, evolution, and evaporation of black holes in the early universe. We then briefly describe in Sec.~\ref{sec:baryon} how baryon number is generated within the context of GUTs. In Sec.~\ref{sec:bhbaryon}, we consider the role that black holes could play in GUT baryogenesis, calculating the magnitude of the baryon asymmetry that results in such scenarios. We then extend this discussion to the case of leptogenesis in Sec.~\ref{sec:lepto}. 
Finally, we discuss and summarize our results in Sec.~\ref{sec:summary}.

\section{Black Holes in the Early Universe}
\label{sec:bh}

It has long been speculated that a significant abundance of black holes may have formed in the early universe~\cite{Carr:1974nx,Carr:1975qj}, either as a consequence of inflation, or during a period involving a phase transition~\cite{Hawking:1982ga,Sasaki:1982fi,Lewicki:2019gmv}. In particular, it has recently been argued that the production of black holes is a relatively generic prediction of simple, single-field inflation models~\cite{Martin:2019nuw,Martin:2020fgl}.

Although the masses of any black holes formed in the early universe is model dependent, it is well motivated to consider values that are less than but comparable to the total energy enclosed within the horizon at the time of their formation~\cite{GarciaBellido:1996qt,Kawasaki:2016pql,Clesse:2016vqa,Kannike:2017bxn,Kawasaki:1997ju,Cai:2018rqf,Yoo:2018esr,Young:2015kda,Clesse:2015wea,Hsu:1990fg,La:1989za,La:1989st,La:1989pn,Weinberg:1989mp,Steinhardt:1990zx,Accetta:1989cr,Holman:1990wq,Hawking:1982ga,Khlopov:1980mg}:
\begin{equation}
\label{horizon}
M_{\rm hor} = \frac{M^2_{\rm Pl}}{2H} \sim 10^{6} \, {\rm g}  \bigg(\frac{10^{13} \, {\rm GeV}}{T}\bigg)^2 \, \bigg(\frac{106.75}{g_{\star}(T)}\bigg)^{1/2},
\end{equation}
where $M_{\rm Pl} = 1.22 \times 10^{19}$ GeV is the Planck mass, $H$ is the Hubble rate, $T$ is the temperature of radiation, and $g_{\star}(T)$ is effective number of relativistic degrees-of-freedom. Thus for reheating temperatures of $T \sim 10^9-10^{13} \, {\rm GeV}$~\cite{Akrami:2018odb,Bezrukov:2011gp,Bezrukov:2007ep}, we should expect $M_{\rm BH} \sim (0.1-1) M_{\rm hor} \sim 10^{5} - 10^{14} \, {\rm g}$.  

Black holes lose mass by radiating particles through the process of Hawking evaporation. In this study, we limit ourselves to the case of Schwarzschild black holes, which evaporate at the following rate~\cite{Hawking:1974sw}:
\begin{eqnarray}
\label{loss}
\frac{dM_{\rm BH}}{dt} &=& -\frac{\mathcal{G} g_{\star, H}(T_{\rm BH}) M^4_{\rm Pl}}{30720 \pi M_{\rm BH}^2}  
\simeq -2.4 \times 10^{15} \, {\rm g \, s}^{-1} \,\bigg(\frac{g_{\star, H}}{316}\bigg) \, \bigg(\frac{10^{6} \, {\rm g}}{M_{\rm BH}}\bigg)^2, 
\end{eqnarray}
where $\mathcal{G} \approx 3.8$ is the appropriate greybody factor, the temperature of a black hole is
\begin{eqnarray}
\label{temp}
T_{\rm BH} = \frac{M^2_{\rm Pl}}{8\pi M_{\rm BH}} \simeq 1.05 \times 10^7 \,{\rm GeV} \, \bigg(\frac{10^{6} \, {\rm g}}{M_{\rm BH}}\bigg),
\end{eqnarray}
%
and the quantity $g_{\star, H}(T_{\rm BH})$ counts the number of particle degrees-of-freedom with masses below $\sim$\,$T_{\rm BH}$, according to the following prescription \cite{1990PhRvD..41.3052M,1991PhRvD..44..376M}:
\be
\label{ghdof}
\hspace{-2cm
}g_{\star,H}(T_{\rm BH}) \equiv \sum_i w_i g_{i,H},   ~~~~~~ g_{i,H}   = 
\begin{cases}
1.82 ~& s = 0 \\
1.0 ~& s = \nicefrac{1}{2} \\
0.41~& s = 1 \\
0.05~ & s = 2
\end{cases}~,
\ee
where $w_i = 2s_i+1$ for massive particles of spin $s_i$, $w_i= 2$ for massless species with $s_i > 0$, and $w_i=1$ for $s_i=0$.
For $T_{\rm BH} \gg 10^2$ GeV ($M_{\rm BH} \ll 10^{11}$ g), the particle content of the Standard Model corresponds to $g_{\star, H} \simeq 108$. Alternatively, the full particle content of the Minimal Supersymmetric Standard Model (MSSM) corresponds to $g_{\star, H} \simeq 316$~\cite{Keith:2020jww}.

It follows from Eq.~(\ref{loss}) that a black hole will evaporate entirely over the following timescale:
\begin{eqnarray}
\label{tevap}
\tau_{\rm BH} = \frac{30720\pi}{\mathcal{G}M^4_{\rm Pl}} \int^{M_{{\rm BH}, i}}_0  \frac{dM_{\rm BH} M_{\rm BH}^2}{g_{\star, H}(T_{\rm BH})}  =
  \frac{ 10240 \pi M_{\rm BH}^3}{ {{\cal G} \langle g_{\star, H}\rangle}  M_{\rm Pl}^4}
\approx 1.4 \times 10^{-10} \, {\rm s} \, \bigg(\frac{M_{{\rm BH},i}}{10^{6} \, {\rm g}}\bigg)^3 \bigg(\frac{316}{\langle g_{\star, H}\rangle}\bigg), 
\end{eqnarray}
where $M_{{\rm BH},i}$ is the initial mass of the black hole and
\be
 \langle g_{\star, H} \rangle^{-1} \equiv \frac{3}{M_{{\rm BH},i}^3} \int^{M_{{\rm BH},i}}_0  \frac{ dM_{\rm BH} M_{\rm BH}^2 }{g_{\star, H}(T_{\rm BH}) }~,
 \ee
 is the value of $g_{\star, H}$ appropriately averaged over the course of the black hole's evaporation. 

As the universe expands, the density of black holes evolves as matter, $\rho_{\rm BH} \propto a^{-3}$, so the fraction of the total energy density in black holes grows proportionally to the scale factor during the era of radiation domination, $\rho_{\rm BH}/\rho_{\rm rad} \propto a^{-3}/a^{-4} = a$~\cite{Lennon:2017tqq,Morrison:2018xla,Hooper:2019gtx}. More quantitatively, the total energy density of the early universe will become dominated by black holes before they finish evaporating if the following condition is satisfied:
\begin{equation}
\label{bhdom}
\frac{\rho_{\rm BH, i}}{\rho_{R, i}} \gsim 4 \times 10^{-9} \, \bigg(\frac{10^{10}\, {\rm GeV}}{T_i}\bigg) \bigg(\frac{10^6 \, {\rm g}}{M_{{\rm BH},i}}\bigg)^{3/2}, 
\end{equation}
where $\rho_{\rm BH, i}$ and $\rho_{R,i}$ represent the energy densities in black holes and radiation at an initial time when the radiation was at a temperature, $T_i$. From this expression, it is clear that black holes could have dominated the total energy density of the early universe prior to their evaporation, even if their initial abundance was quite small. 

If the energy density of the universe was dominated by black holes at the time of their evaporation, the Hawking radiation would fill the universe with a hot bath of radiation, reheating the universe to a temperature, $T_{\rm RH}$. Under the approximation of instantaneous evaporation at $t = \tau_{\rm BH}$, the temperature of this radiation can be calculated as follows:
\be
\label{TRH1}
\rho_{\rm BH} = \frac{   3 M_{\rm Pl}^2 H(\tau_{\rm BH})^2 }{   8\pi       }    =  
\frac{    M_{\rm PL}^2 }{   6\pi   \tau_{\rm BH}^2    }        \approx
\frac{   \pi^2 g_\star(T_{\rm RH})     }{30 }T_{\rm RH}^4  ~.
\ee 
Applying \Eq{tevap} in the limit where $g_{\star,H}$ is constant, and solving for $T_{\rm RH}$ in \Eq{TRH1}, we obtain 
\begin{eqnarray}
\label{TRH-BH}
T_{\rm RH} \simeq 50 \, {\rm GeV} \, \bigg(\frac{10^6 \, {\rm g}}{M_{{\rm BH},i}}\bigg)^{3/2}  \bigg(\frac{g_{\star, H}(T_{\rm BH})}{316}\bigg)^{1/2}  \bigg(\frac{90}{g_{\star}(T_{\rm RH})}\bigg)^{1/4}~,
\end{eqnarray}
which should not be confused with the temperature of reheating following inflation. Throughout this paper, we will use $T_{\rm RH}$ to denote the temperature of the universe following the evaporation of a black hole population, and will focus on black holes which evaporate before the onset of BBN ($M_{{\rm BH}, i} \lsim 5 \times 10^8 \, {\rm g}$)~\cite{Keith:2020jww,Carr:2009jm}.

\section{The Generation of Baryon Number in Grand Unified Theories}
\label{sec:baryon}

A generic feature of GUTs are heavy bosons with baryon number violating couplings. For example, in $SU(5)$ there are a total of 12 additional gauge bosons (denoted by $X$ and $Y$), each of which forms an $SU(3)_C$ triplet and $SU(2)_L$ doublet. The interactions of these particles connect quarks and leptons in a way that violates the conservation of baryon and lepton number, leading to processes such as $p \rightarrow \pi^0 e^+$. Although the precise rate at which GUT gauge bosons mediate proton decay is somewhat model dependent, a reasonable estimate is given by: 
\begin{eqnarray}
\Gamma_p \approx \frac{g^4_{\rm GUT}}{16\pi^2} \frac{m_p^5}{M^4_{X,Y}},
\end{eqnarray}
where $g_{\rm GUT}$ is the strength of the gauge coupling of the GUT force, and $M_{X,Y}$ are the masses of the gauge bosons mediating the decay. For $g_{\rm GUT} \approx 1/25$ (the value at which the Standard Model gauge couplings unify in supersymmetric GUTs), the current constraint on proton decay from Super-Kamiokande of $\tau_{p}/{\rm Br}(p \rightarrow \pi^0 e^+) < 1.67 \times 10^{34}\,{\rm years}$~\cite{Miura:2016krn,Miura:2016krn} allows us to restrict $M_{X,Y} \gsim 3\times 10^{15}\, {\rm GeV}$. Given that gauge coupling unification takes place at $\sim$\,$10^{16}$ GeV in supersymmetric $SU(5)$, this class of models remains consistent with existing constraints on proton decay.

In addition to the Higgs bosons that break the gauge symmetry of the GUT (which are contained in the adjoint representation of the GUT group), additional Higgs bosons are required to break electroweak symmetry at the weak scale, and to give masses to the Standard Model fermions. In $SU(5)$, the minimal representation that can accommodate this is a $\bold{5}_H$ or $\bold{\bar{5}}_H$ (in supersymmetric GUTs, both $\bold{5}_H$ and $\bold{\bar{5}}_H$ Higgs multiplets are required in order to cancel anomalies)~\cite{Dimopoulos:1981yj,Dimopoulos:1981zb,Ibanez:1981yh,Ibanez:1981yh,Einhorn:1981sx,Marciano:1981un}. Each $\bold{5}_H$ or $\bold{\bar{5}}_H$ multiplet consists of an $SU(2)$ doublet (such as that corresponding to a Higgs doublet near the electroweak scale) and a color triplet. Unlike the doublets, the Higgs triplets violate the conservation of both baryon and lepton number by coupling to the following combinations of Standard Model fermions (in each generation): $u e^-$, $d\nu$, and $ud$~\cite{Harvey:1981yk}. Similar to those of the GUT gauge bosons, the interactions of the Higgs triplets lead to the following rate of proton decay:
\begin{eqnarray}
\label{gammap}
\Gamma_p \sim y_{u,d}^4 \frac{m_p^5}{M^4_{\cal T}},
\end{eqnarray}
where $M_{\cal T}$ is the mass of the Higgs triplet, and $y_{u,d}$ are its baryon and lepton number violating Yukawa couplings to first generation fermions. If the Higgs triplet has Yukawa couplings that are similar in magnitude to those of the electroweak Higgs doublet ($y_{u,d} \sim \mathcal{O}(m_u/m_t)$), constraints from proton decay require that $M_{\cal T} \gsim 3 \times 10^{11} \, {\rm GeV}$.


In conventional GUT baryogenesis, the baryon asymmetry of the universe is generated through the baryon number violating (and C and CP violating) decays of $X$, $Y$, and ${\cal T}$ particles. The requirement that these particles be heavy enough to not induce an unacceptable rate of proton decay provides a challenge for this scenario, as the production of a large abundance of such particles in the early universe would appear to require a rather high reheating temperature. In the following section, we will consider an alternative scenario, in which the GUT bosons that generate the baryon asymmetry are produced at somewhat later times, through the Hawking evaporation of primordial black holes. 

\section{The Role of Black Holes in GUT Baryogenesis}
\label{sec:bhbaryon}

For black holes in the mass range under consideration in this study ($M_{{\rm BH}, i} \sim 10^0-10^9 \, {\rm g}$), their initial temperatures ($T_{\rm BH} \sim 10^4-10^{13}\,{\rm GeV}$) are generally well below those required to radiate particles with GUT-scale masses. Consequently, Hawking evaporation only becomes capable of producing such massive particles once the black hole has already lost most of its mass, causing its temperature to become comparable to or greater than the mass of the particles in question. Here, we will focus on the Hawking radiation of Higgs triplets, which is efficient when $T_{\rm BH} \gsim M_{\cal T}$.  In this section, we calculate the quantities and average energies of the heavy particles produced through Hawking evaporation and use these results to obtain the baryon asymmetry that is generated when these particles decay in accordance with the Sakharov conditions. 
For future convenience, we define a the parameter 
\begin{equation}
\epsilon_{\cal T} \equiv \sum_f B_f \frac{ \Gamma_{\cal T}({\cal T} \rightarrow f)-\Gamma_{\cal T}(\bar{{\cal T}}\rightarrow \bar{f})}{\Gamma_{\cal T}},
\end{equation}
which represents the average baryon yield per triplet decay in the limit where washout effects can be neglected; all of our results below will be proportional to this value. 
Here, the sum is over all final states, $f$, with net baryon number $B_f$, and $\Gamma_{\cal T}$ is the total width of the Higgs triplet.  Note that in order for $\epsilon_{\cal T} \ne 0$, one generally requires a second baryon number violating boson to be present in the loop-level decay diagrams. In the case of supersymmetric $SU(5)$, this is naturally accommodated by the two Higgs Triplets that are contained within the $\bold{5}_H$ and $\bold{\bar{5}}_H$ representations.

\subsection{The Kinematics of Black Hole Particle Production}

Approximating that the average radiated triplet has an energy, $E_{\cal T} \approx 3 T_{\rm BH}$, we can estimate the number of Higgs triplets that are produced over the course of a black hole's lifetime:
\begin{eqnarray}
\label{NT1}
N_{\cal T} \sim \frac{M_{\rm Pl}^2}{24\pi} \int^{M_{\rm Pl}}_{M_{\cal T}} \frac{dT_{\rm BH}}{T_{\rm BH}^3} \, \bigg(\frac{g^{\cal T}_{H}}{g_{\star, H}}\bigg) 
\simeq 6 \times 10^{10} \, \bigg(\frac{10^{12}\,{\rm GeV}}{M_{\cal T}}\bigg)^2 \, \bigg(\frac{g^{\cal T}_{H}/g_{\star, H}}{0.065}\bigg),
\end{eqnarray}
where  $g^{\cal T}_{H} \approx 6 \times 1.82 \approx 10.9$ is the Hawking radiation weight per Higgs triplet (corresponding to 21.8 for the $\bold{5}_H$ and $\bold{\bar{5}}_H$ Higgs representations required in supersymmetric GUTs), and $g_{\star, H} \simeq 316 + g^{\cal T}_{H}$ for the low-energy particle content associated with the MSSM. This result assumes that Higgs triplet production begins instantaneously (and without suppression) at $T_{\rm BH} = M_{\cal T}$. 

A more careful result can be obtained by integrating over a Bose-Einstein distribution of radiated particles at temperature $T_{\rm BH}$. The rate of $\cal T$ production per unit black hole area can be written as
\be
\frac{dN_{\cal T}}{ dt \, dA} = \frac{ {\cal G} g^{\cal T}_{H}}{4} \int \frac{d^3p}{(2\pi)^3} \frac{1}{e^{E/T_{\rm BH}} - 1},
\ee
where ${\cal G}$ is a greybody factor and the factor of $1/4$ converts particle production per volume to production per area \cite{Hook:2014mla}. Integrating over the surface area, and applying Eqs.~(\ref{loss}) and~(\ref{temp}), the total rate of $\cal T$ production becomes
\be
\label{rate-per-area}
\frac{dN_{\cal T}}{ dt }
  =  \pi r_s^2 {\cal G} g^{\cal T}_{H} \int \frac{d^3p}{(2\pi)^3} \frac{1}{e^{E/T_{\rm BH}} - 1},
\ee
where $r_s  = 2 M_{\rm BH}/M^2_{\rm Pl} = (4 \pi T_{\rm BH})^{-1}$ is the Schwarzschild radius, 
so the total number of triplets produced per black hole is  
\be
\label{NT}
N_{\cal T}  =
 \frac{ 15  M^2_{\rm Pl}}{ 4 \pi^3    }
\int^{M_{\rm Pl}}_{T_{{\rm BH,}i}} \frac{  dT_{\rm BH}  }{T^6_{\rm BH}}
\left( \frac{ g^{\cal T}_{H}}{ g_{\star, H}} \right)
 \int \frac{d^3p}{(2\pi)^3} \frac{1}{e^{E/T_{\rm BH}} - 1}.
\ee
In the limit of $T_{\rm BH, i} \ll M_{\cal T}$, this reduces to
\be
N_{\cal T} \approx 5.0 \times 10^{10} \, \bigg(\frac{10^{12}\,{\rm GeV}}{M_{\cal T}}\bigg)^2 \, \bigg(\frac{g^{\cal T}_{H}/g_{\star, H}}{0.065}\bigg),
\ee
while in the opposite limit ($T_{\rm BH, i} \gg M_{\cal T}$), we obtain the following:
\be
N_{\cal T}   \approx  5.7 \times 10^{13} \, \bigg(\frac{M_{\rm BH, i}}{10^{2}\,{\rm g}}\bigg)^2 \, \bigg(\frac{g^{\cal T}_{H}/g_{\star, H}}{0.065}\bigg).
\ee
Similarly, the mass loss rate to $\cal T$ emission is
\be
\frac{dM_{\rm BH \to \cal T}}{dt}   =
 \pi r_s^2 {\cal G} g^{\cal T}_{H} 
\int \frac{d^3p}{(2\pi)^3} \frac{E}{e^{E/T_{\rm BH}} - 1},
\ee
so the total mass lost to $\cal T$ emission is
\be
\Delta M_{\rm BH \to \cal T}  =
 \frac{ 15  M^2_{\rm Pl}}{ 4 \pi^3    }
\int^{M_{\rm Pl}}_{T_{{\rm BH,}i}} \frac{  dT_{\rm BH}  }{T^6_{\rm BH}}
\left( \frac{ g^{\cal T}_{H}}{ g_{\star, H}} \right)
 \int \frac{d^3p}{(2\pi)^3} \frac{E}{e^{E/T_{\rm BH}} - 1} ~,
\ee  
and the average Lorentz boost per $\cal T$ particle weighted over the full BH lifetime is 
\be
\langle \gamma \rangle_{\rm BH}^{\cal T} = 
\frac{    \Delta M_{\rm BH \to \cal T} }{  N_{\cal T}  M_{\cal T}   } =
  \begin{cases}
2.3~~~~~~~~~~~~~~ \, \,\,\,\,\,\,\,\,\,\,\,\,\,\,\,\,\,\,\,\,\,\,\,\,\,\,\,\,\,\, M_{\cal T} \gg T_{{\rm BH}, i} \\
5.4 \times (  T_{{\rm BH}, i} /  M_{\cal T}  ) ~~~~~~~~~~\,  M_{\cal T} \ll T_{{\rm BH}, i}\\
\end{cases}~~   .
 \ee
 For relatively small values of $M_{\cal T}$ and/or $M_{\rm BH}$, the average Lorentz factor could potentially be large, significantly delaying the decays of the Higgs triplets through time dilation. This will be the case, however, only if the Higgs triplets do not 
efficiently transfer their kinetic energy to the thermal bath though elastic scattering. The cross section for Triplet scattering with a gluon is given by $\sigma_{{\cal T}g \to {\cal T}g}  \sim \alpha^2_s T^2/M^4_{\cal T}$, which is sufficiently large to establish kinetic equilibrium if the scattering rate exceeds the rate of Hubble expansion:
\be
\frac{\sigma_{{\cal T}g \to {\cal T}g}  \, n_g  }{H}  \sim 
\frac{1.2  \alpha_s^2  T_{\rm RH}^3  \mpl  }{\sqrt{g_\star} M_{\cal T}^4      } \gsim 1 ~~\implies 
M_{\cal T} \lsim 2 \times 10^5 \, {\rm GeV} \left(          \frac{ 10^6\,{\rm g}    }{   M_{{\rm BH},i}   }     \right)^{9/8},
\ee
where $n_g$ is the number density of gluons in the thermal bath and, in expression on the right, we have evaluated at the reheating temperature following an era of black hole domination, $T_{\rm RH}$. Across most of the parameter space in which the Higgs triplets are produced with large Lorentz factors, these particles are rapidly cooled through scattering with gluons, efficiently preventing any significant time dilation of their decays.

\subsection{Baryogenesis From a Black Hole Dominated Era}

In calculating the magnitude of the baryon asymmetry that is produced in the decays of the Higgs triplets radiated from black holes, we will first consider the case in which black holes dominated the energy density of the universe prior to their evaporation. Such an initial condition can be realized for any initial black hole energy density that satisfies the 
inequality in \Eq{bhdom} and, once realized, is insensitive to any prior state of the early universe. 

We begin by determining the number density of black holes at the time of their evaporation. From the definition of the Hubble rate in matter domination, we have 
\be
H(\tau_{\rm BH})^2 = \frac{8\pi}{3} \frac{\rho_{\rm BH}}{M_{\rm Pl}^2} =   \frac{4}{9\tau_{\rm BH}^2} ,~~
\ee
where $\tau_{\rm BH}$ is  given in \Eq{tevap}. Note that the number density of black holes, $n_{\rm BH} = \rho_{\rm BH}/M_{\rm BH}$, at the time of evaporation is uniquely specified by the mass of the black holes.
From this, the resulting baryon asymmetry from GUT Higgs triplet decays is 
\begin{eqnarray}
\label{BHdom}
Y_B &=& \frac{n_{\rm BH}}{s} \, \epsilon_{\cal T} N_{\cal T} = \frac{3 g_{\star}(T_{\rm RH}) T_{\rm RH}}{4 g_{\star S} (T_{\rm RH}) M_{{\rm BH},i}}  \, \epsilon_{\cal T} N_{\cal T} ,
\ee
where $s$ is the entropy density at the time of evaporation, $g_{\star S}$ is the number of degrees-of-freedom in entropy, and 
$N_{\cal T}$ from \Eq{NT} is the number of triplets produced via Hawking radiation over the lifetime of a single 
black hole. Importantly, here $T_{\rm RH}$ is the Standard Model radiation temperature that results entirely from Hawking
radiation according to \Eq{TRH-BH}.
Assuming triplets decay promptly, the baryon yield in the $T_{\rm BH, i} \ll M_{\cal T}$ limit is given by:
\be
\label{BHdom2a}
Y_B \approx  3.6 \times 10^{-10}  \,  
\bigg(\frac{g_{\star}(T_{\rm RH})}{g_{\star S}(T_{\rm RH})}\bigg)
 \bigg(\frac{10^{12}\,{\rm GeV}}{M_{\cal T}}\bigg)^2 \, \bigg(\frac{10^2 \, {\rm g}}{M_{{\rm BH},i} }\bigg)^{5/2}  \bigg(\frac{\epsilon_{\cal T}}{10^{-2}}\biggr)  \bigg(\frac{g^{\cal T}_{H}/g_{\star, H}}{0.065}\bigg)   \bigg(\frac{g_{\star H}}{316}\bigg)^{1/2}
 \bigg(\frac{90}{g_{\star}(T_{\rm RH})}\bigg)^{1/4}.
\ee
For a fixed $Y_B$, this relation which predicts a $M_{\cal T} \propto M_{\rm BH}^{5/4}$ relationship,
which corresponds to the scaling of the dashed contours in Fig. \ref{fig1} for large values of $M_{\cal T}$ in the upper left. 
In the $T_{\rm BH, i} \gg M_{\cal T}$ limit, we instead arrive at:
\be
\label{BHdom2b}
Y_B \approx  3.2 \times 10^{-11}  \,  
\bigg(\frac{g_{\star}(T_{\rm RH})}{g_{\star S}(T_{\rm RH})}\bigg) \, \bigg(\frac{10^8 \, {\rm g}}{M_{{\rm BH},i} }\bigg)^{1/2}  \bigg(\frac{\epsilon_{\cal T}}{10^{-2}}\biggr)  \bigg(\frac{g^{\cal T}_{H}/g_{\star, H}}{0.065}\bigg)   \bigg(\frac{g_{\star H}}{316}\bigg)^{1/2}
 \bigg(\frac{90}{g_{\star}(T_{\rm RH})}\bigg)^{1/4},
\ee
which is independent of $M_{\cal T}$ 
and corresponds to the vertical portions of the dashed contours in Fig.~\ref{fig1}.

\begin{figure}[t]
\includegraphics[width=0.55\textwidth]{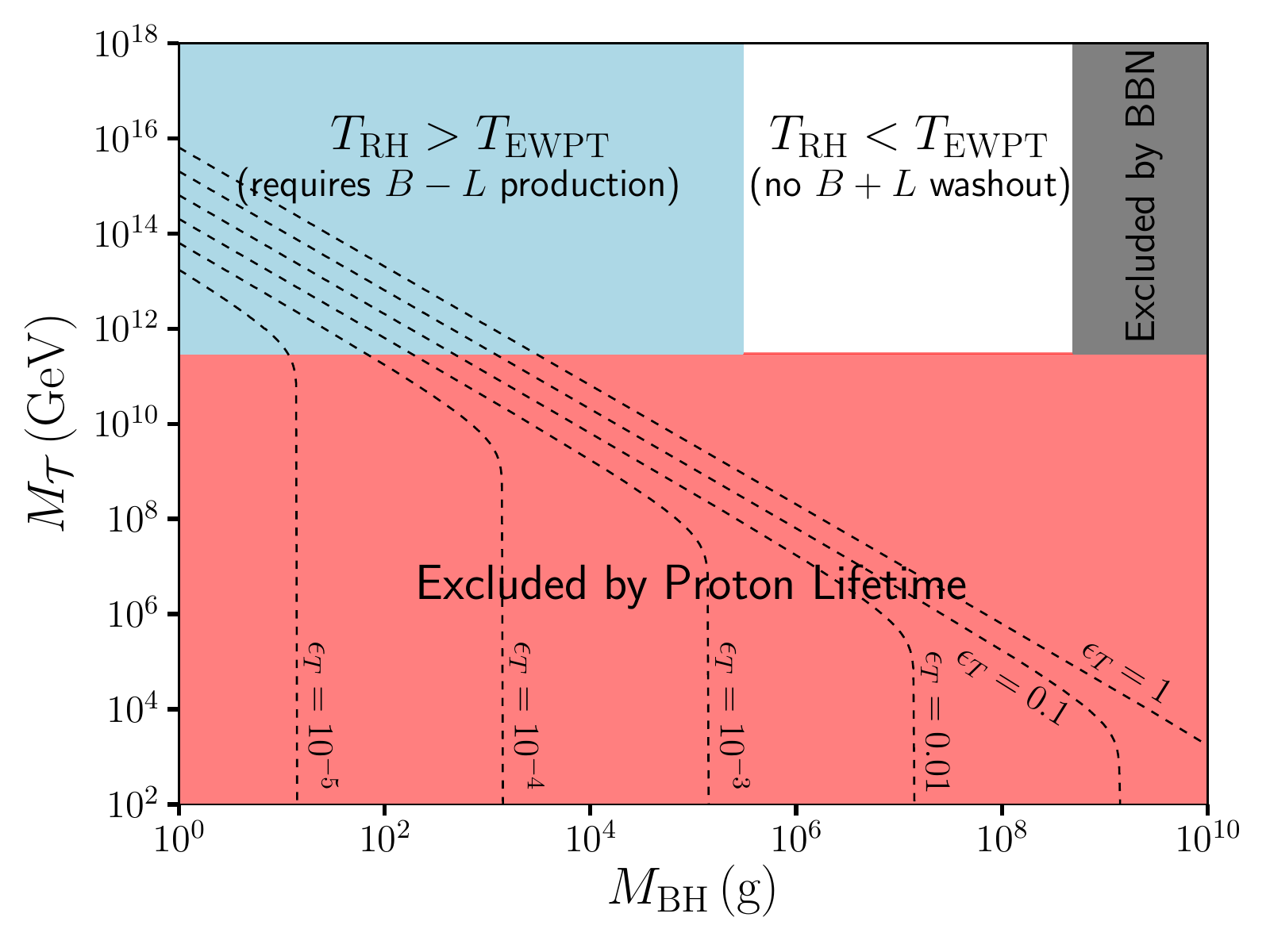} 
\caption{Parameter space for producing the observed baryon asymmetry as a function
of $M_{\cal T}$ and $M_{\rm BH}$. The dashed curves represent regions where $Y_B = 8.8 \times 10^{-11}$, for various values of $\epsilon_{\cal T}$. In this figure, we have assumed that black holes dominated the energy density of the universe prior to their evaporation, and have taken a common value for Yukawa couplings of the Higgs (electroweak) doublet and (color) triplet. In the region not excluded by proton decay constraints, the universe is reheated by black hole evaporation to a temperature greater than that of the electroweak phase transition, enabling sphalerons to wash out any net value of $B+L$ that was generated through the decays of Higgs triplets. To avoid sphaleron washout, one could consider GUTs that are not based on $SU(5)$, which in some cases can be capable of generating a net $B-L$, which remains conserved even at temperatures above the electroweak phase transition.}
\label{fig1}
\end{figure}

If $M_{\cal T} \gsim 3 \times 10^{11} \, {\rm GeV}$ (as required by proton decay constraints, assuming that the Yukawa couplings of the Higgs triplet are equal in magnitude to those of the Higgs doublet), obtaining the measured baryon asymmetry requires \cite{Barrow:1990he,Baumann:2007yr,Morrison:2018xla}
\be
\label{BHdom3}
 Y_B  \approx 8.8 \times 10^{-11} \implies M_{{\rm BH},i} \lsim 5 \times 10^2 \, {\rm g} \, \left( \frac{ \epsilon_{\cal T}}{10^{-2}  } \right)^{2/5}~.
 \ee
However, from \Eq{TRH-BH}, for black holes in this mass range, the process of Hawking evaporation ends when the temperature of the universe is well above that of the electroweak phase transition, enabling sphalerons to wash out any net value of $B+L$ that was generated through the Higgs triplet decays. To avoid sphaleron washout, one could consider GUTs that are not based on $SU(5)$ and that are capable of generating a net abundance of $B-L$, which remains conserved even at temperatures above the electroweak phase transition. This situation is summarized in Fig.~\ref{fig1}.

\subsection{Baryogenesis From a Subdominant Black Hole Population}

One can obtain a more general solution for $Y_B$ by rescaling the value of $\epsilon_{\cal T}$ in Eqs.~(\ref{BHdom2a}) and~(\ref{BHdom2b}) by $\rho_{\rm BH}/\rho_{\rm tot}$, as evaluated at the time of evaporation.  Unlike in those scenarios which included a black hole dominated era, the result in this case depends on the initial conditions and, more specifically, on the composition of the universe at the time of black hole formation. 

We can write the baryon yield in this case as follows:
\be
\label{fracyield}
Y_B \equiv \frac{                   n_{{\rm BH}}           }{     s}  \epsilon_{\cal T} N_{\cal T} = \frac{\rho_{\rm BH}}{\rho_{\rm tot}}\bigg|_{\tau} 
\frac{      3   g_\star(T_{\tau}  )          }{4 g_{\star, S}(T_{\tau}  )  }
\frac{      T_{\tau}    }{M_{{\rm BH}, i}}     \epsilon_{\cal T}   N_{\cal T} ,
\ee
where $T_{\tau}$ is the temperature of the radiation bath and $\rho_{\rm BH}/\rho_{\rm tot} |_{\tau}$ represents the ratio of these energy densities, each evaluated at the time of evaporation. As the universe expands, the fraction of the energy density in black holes grows linearly with the scale factor, so at evaporation we have
\be
 \frac{\rho_{\rm BH}}{\rho_{\rm tot}}\bigg|_{\tau}  =  \frac{\rho_{\rm BH}}{\rho_{\rm tot}}\bigg|_{i} \, \times \,\bigg(\frac{T_i}{T_{\tau}}\bigg), 
%
\ee
where $T_i$ is the temperature of the radiation bath  at the time of black hole formation and the temperature at 
evaporation is determined by the condition $H = (2\tau_{\rm BH})^{-1}$, which yields
\be
 T_\tau = 
\left(      \frac{M_{\rm Pl}}{3.33 \sqrt{g_\star} \tau_{\rm BH}}  \right)^{1/2}~.
\ee

In a case with no black hole dominated era, once again assuming that the triplets decay promptly, the baryon yield in the $T_{\rm BH, i} \ll M_{\cal T}$ limit is given by:
\be
Y_B \approx  8.5 \times 10^{-20}  \,  
\bigg(\frac{g_{\star}(T_{\rm RH})}{g_{\star S}(T_{\rm RH})}\bigg)
 \bigg(\frac{10^{12}\,{\rm GeV}}{M_{\cal T}}\bigg)^2 \, \bigg(\frac{10^2 \, {\rm g}}{M_{{\rm BH},i} }\bigg)  \bigg(\frac{\epsilon_{\cal T}}{10^{-2}}\biggr)  \bigg(\frac{g^{\cal T}_{H}/g_{\star, H}}{0.065}\bigg) \, \bigg( \frac{\rho_{\rm BH}/\rho_{\rm tot} |_{i} }{10^{-12}}\bigg) \, \bigg(\frac{T_i}{10^{10} \, {\rm GeV}}\bigg).
\ee
Whereas in the $T_{\rm BH, i} \gg M_{\cal T}$ limit, we instead arrive at
\be
Y_B \approx  7.5 \times 10^{-12}  \,  
\bigg(\frac{g_{\star}(T_{\rm RH})}{g_{\star S}(T_{\rm RH})}\bigg) \, \bigg(\frac{M_{{\rm BH},i} }{10^8 \, {\rm g}}\bigg)  \bigg(\frac{\epsilon_{\cal T}}{10^{-2}}\biggr)  \bigg(\frac{g^{\cal T}_{H}/g_{\star, H}}{0.065}\bigg)  \,  \bigg( \frac{\rho_{\rm BH}/\rho_{\rm tot} |_{i} }{10^{-12}}\bigg) \, \bigg(\frac{T_i}{10^{10} \, {\rm GeV}}\bigg), 
 \ee
which is independent of $M_{\cal T}$ as in the corresponding regime during black hole domination in \Eq{BHdom2b}.

\subsection{Reducing The Yukawa Couplings of the Higgs Triplet}

So far in this study, we have found that the observed baryon asymmetry, $Y_B \approx 8.8 \times 10^{-11}$, can be produced through the Hawking evaporation of baryon number violating Higgs triplets without inducing an unacceptable rate of proton decay only in scenarios with relatively light black holes, $M_{\rm BH} \lsim 5 \times 10^2 \,{\rm g}$. Black holes in this mass range evaporate leaving the universe at a temperature well above the electroweak phase transition, enabling sphalerons to wash out any net value of $B+L$ that may have been previously generated. One possible solution to this problem would be to consider GUTs which allow for the production of net $B-L$, which remains conserved at all temperatures. In this section, we will consider an alternative solution which involves reducing the Yukawa couplings of the Higgs triplet.

In the previous subsections, we assumed that the Yukawa couplings of the Higgs triplet are equal or similar in magnitude to those of the electroweak Higgs doublet. Given that the $SU(3)$ triplet and $SU(2)$ doublet are each part of the same $\bold{5}_H$ or $\bold{\bar{5}}_H$ multiplet, this is a reasonable and relatively minimal assumption. That being said, the large hierarchy between the mass of the Higgs triplet (as required by constraints from proton decay to be $M_{\cal T} \gsim 3 \times 10^{11} \, {\rm GeV}$) and the much smaller mass of the electroweak Higgs doublet would appear to involve an unacceptable degree of fine tuning, and this has motivated a great deal of model buliding, featuring a variety of potential solutions~\cite{Witten:1981kv,Masiero:1982fe,Grinstein:1982um,Dimopoulos:1981xm,Dimopoulos:1981xm,Inoue:1985cw,Inoue:1985cw,Hisano:1994fn,Altarelli:2000fu,Babu:1993we,Babu:1994dq,Babu:2010ej,Babu:1994kb,Lucas:1995ic,Blazek:1996yv,Barr:1997hq,Babu:1998wi,Dermisek:2000hr,Barbieri:1993wz,Barbieri:1992yy,Berezhiani:1995sb,Shafi:2001iu,Dvali:1996sr,Dvali:1992hc,Dvali:1995hp,Rakshit:2003wj,Haba:2002if,Goldberger:2002pc}. Motivated by such considerations, we examine in this subsection scenarios in which the Yukawa couplings of the Higgs triplet are independent from those of the electroweak doublet, allowing us to consider much lower values of $M_{\cal T}$ without inducing an unacceptable rate of proton decay. By suppressing the Yukawa couplings of the Higgs triplet, we find that it is possible to generate the observed baryon asymmetry from more massive black holes, which evaporate leaving the universe at a temperature below the electroweak phase transition, naturally avoiding the problem of sphaleron washout. 

In the case of relatively light triplets, their annihilation rate can potentially exceed the rate of Hubble expansion, leading to the depletion of the triplet abundance before their decays can generate an appreciable baryon asymmetry. In the low-velocity limit, the cross section for triplet annihilation to gluons is given by the following:
\begin{equation}
\sigma_{{\cal TT}\rightarrow gg} v \simeq \frac{7 \pi \alpha_s^2}{18 M^2_{\cal T}}.
\end{equation}

The annihilation rate is the product of this cross section and the number density of triplets immediately after black hole evaporation, $\Gamma_{{\cal TT}\rightarrow gg} = \sigma_{{\cal TT}\rightarrow gg} v \, n_T = \sigma_{{\cal TT}\rightarrow gg} v \, N_{\cal T} \rho(T_{\rm BH})/M_{\rm BH}$. Comparing this to the Hubble rate during radiation domination at a temperature $T_{\rm RH}$, we arrive at the condition for efficient triplet annihilation (evaluated in the $T_{{\rm BH},i} \gg M_{\cal T}$ limit):
\begin{eqnarray}
M_{\cal T} \lsim 3 \times 10^6 \, {\rm GeV} \times \bigg(\frac{10^6 \, {\rm g}}{M_{\rm BH}}\bigg).
\end{eqnarray}

The parameter space in this class of scenarios is summarized in Fig.~\ref{fig2}, for the choice of Yukawa couplings which saturate the bound from proton decay. Throughout this study, whenever we alter the Yukawa couplings of the Higgs triplet, we simply rescale these quantities and parameterize this change in terms of their values relative to the constraints derived from proton decay. More specifically, we define
\begin{equation}
y_{\rm max} = y_{\rm DT} \times \bigg(\frac{M_{\cal T}}{3\times 10^{11} \, {\rm GeV}}\bigg),
\end{equation}
where $y_{DT}$ denotes the values of the Yukawa couplings in the traditional case of doublet-triplet unification. In the following subsection, we will consider the case in which the Yukawas are even smaller ($y \ll y_{\rm max}$), leading to scenarios featuring long-lived Higgs triplets.

\begin{figure}[t]
\includegraphics[width=0.55\textwidth]{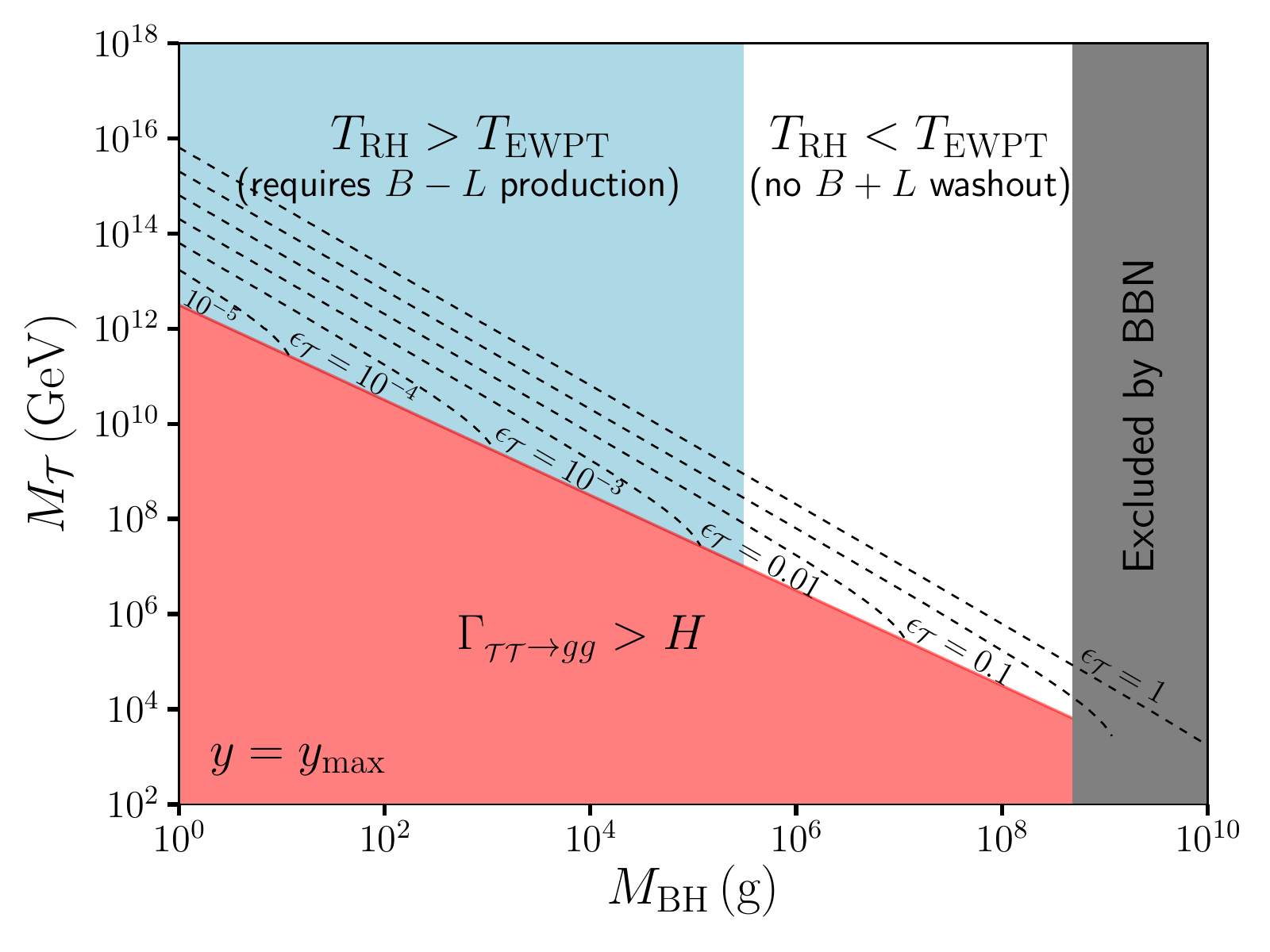} 
\caption{As in Fig.~\ref{fig1}, but rescaling the Yukawa couplings of the Higgs triplet such that they saturate the constraint from proton decay. By suppressing the Yukawa couplings of the Higgs triplet, we find that it is possible to generate the observed baryon asymmetry with black holes of much larger mass, which evaporate after the electroweak phase transition. In the blue region, the triplets decay prior to the electroweak phase transition, enabling sphalerons to wash out any net value of $B+L$ that was generated through the Higgs triplet decays (the baryon asymmetry can only be generated in this parameter space if one considered GUTs which are capable of generating a net value of $B-L$). In the white region, any baryon asymmetry that is produced through triplet decays is protected from the effects of sphaleron washout. In the red region, the annihilation rate of the triplets exceeds that of Hubble expansion, leading to the depletion of the triplet abundance before an appreciable baryon asymmetry can be generated. Throughout this figure, we have assumed that black holes dominate the energy density of the universe prior to their evaporation.}
\label{fig2}
\end{figure}

\begin{figure}[t] 
\includegraphics[width=0.55\textwidth]{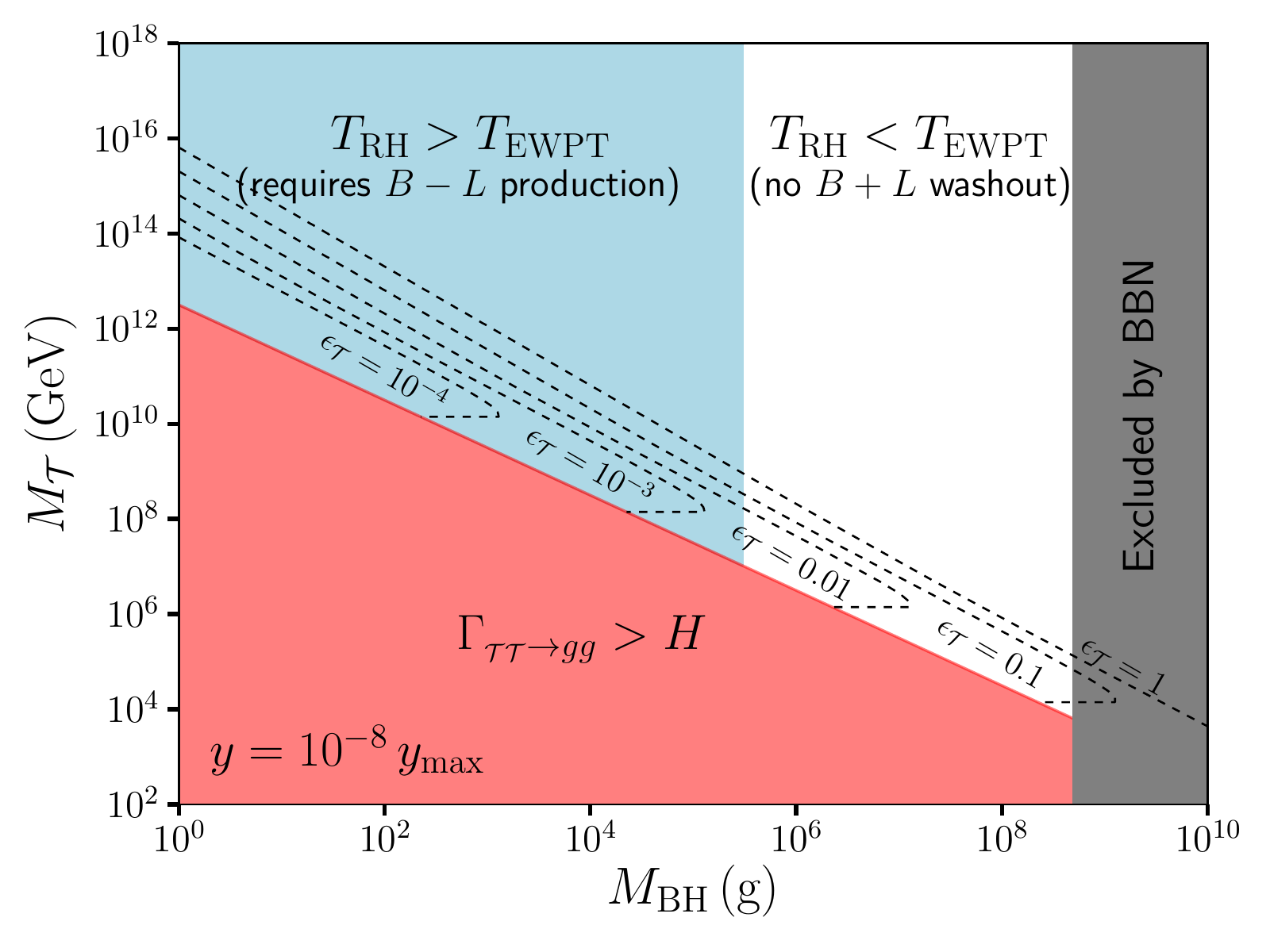} 
\caption{As in Fig.~\ref{fig2}, but setting the Yukawa couplings of the Higgs triplet to $10^{-8}$ times the maximum value consistent with the constraints from proton decay. By suppressing these Yukawa couplings, the triplets become long-lived, leading to an era in which the triplets dominate the energy density of the universe prior to their decay (see Appendix~\ref{A1}). While the diagonal contours of constant $Y_B$ are naively similar to those shown in Fig.~\ref{fig2}, the underlying physics is quite different (see Appendix~\ref{moderate}). Furthermore, where the contours of constant $Y_B = 8.8 \times 10^{-11}$ extend horizontally, the triplets do not only dominate the energy density, but dominate it by a factor greater than $Y_B^{-1}$, leading to the generation of a baryon asymmetry that is independent of the mass of the black holes (see Appendix~\ref{way-out}). 
}
\label{fig2LL}
\end{figure}

\subsection{Long-Lived Triplets}
\label{appA}

Up to this point in the discussion, we have treated the decays of the Higgs triplets as instantaneous. This is a good approximation so long as the decay rate is rapid compared to the rate of Hubble expansion at the time. The width of the Higgs triplet is approximately given as \cite{Harvey:1981yk}: 
\begin{eqnarray}
\label{Twidth}
\Gamma_{\cal T} \approx \frac{M_{\cal T} (4 y_t^2+3y_b^2)}{16 \pi},
\end{eqnarray}
where $y_{t,b}$ are the Yukawa couplings of the Higgs triplet to third generation quarks. If we set the magnitude of these Yukawa couplings such that they saturate the bounds from proton decay, $y_{t,b} \lsim M_{\cal T}/(3\times 10^{11} \, {\rm GeV})$, we arrive at:
\begin{eqnarray}
\Gamma_{\cal T} \lsim 1.6 \times 10^{-15} \, {\rm GeV} \, \bigg(\frac{M_{\cal T}}{\rm TeV}\bigg)^3.
\end{eqnarray}
%


In the period of time after the completion of black hole evaporation, but before the decay of the triplets, the fraction of the total energy density consisting of non-relativistic triplets will increase by a factor of $(a_{\rm decay}/a_{\rm evap}) \approx (\tau_{\cal T}/\tau_{\rm BH})^{1/2}$. Furthermore, if $\tau_{\cal T} \gsim \tau_{\rm BH} \times (g_{\star, H}/g^T_{\star, H})^2$, the Higgs triplet population will dominate the total energy density of the universe before they decay, potentially altering the results of our calculation. 
%
For example, if we take the Yukawa couplings of the triplet to be eight orders of magnitude smaller than the maximum value consistent with the constraints from proton decay, $y=10^{-8} \, y_{\rm max}$, a triplet-dominated era will occur if $M_{\cal T} \lsim 10^8 \, {\rm GeV} \times (10^6 \, {\rm g}/M_{\rm BH})$. This case is illustrated in Fig.~\ref{fig2LL}, where the contours of constant $Y_B = 8.8 \times 10^{-11}$ extend diagonally to the lower-right, until $M_{\cal T} \simeq 1.4 \times 10^2 \times \epsilon_{\cal T}^{-2}$, at which point we enter a regime in which the triplets do not only dominate the energy density, but dominate it by a factor greater than $Y_B^{-1}$. In this later limit, the resulting baryon asymmetry is independent of the mass of the black holes.




\section{Application to Leptogenesis}
\label{sec:lepto}

\begin{figure}[t]
\includegraphics[width=0.55\textwidth]{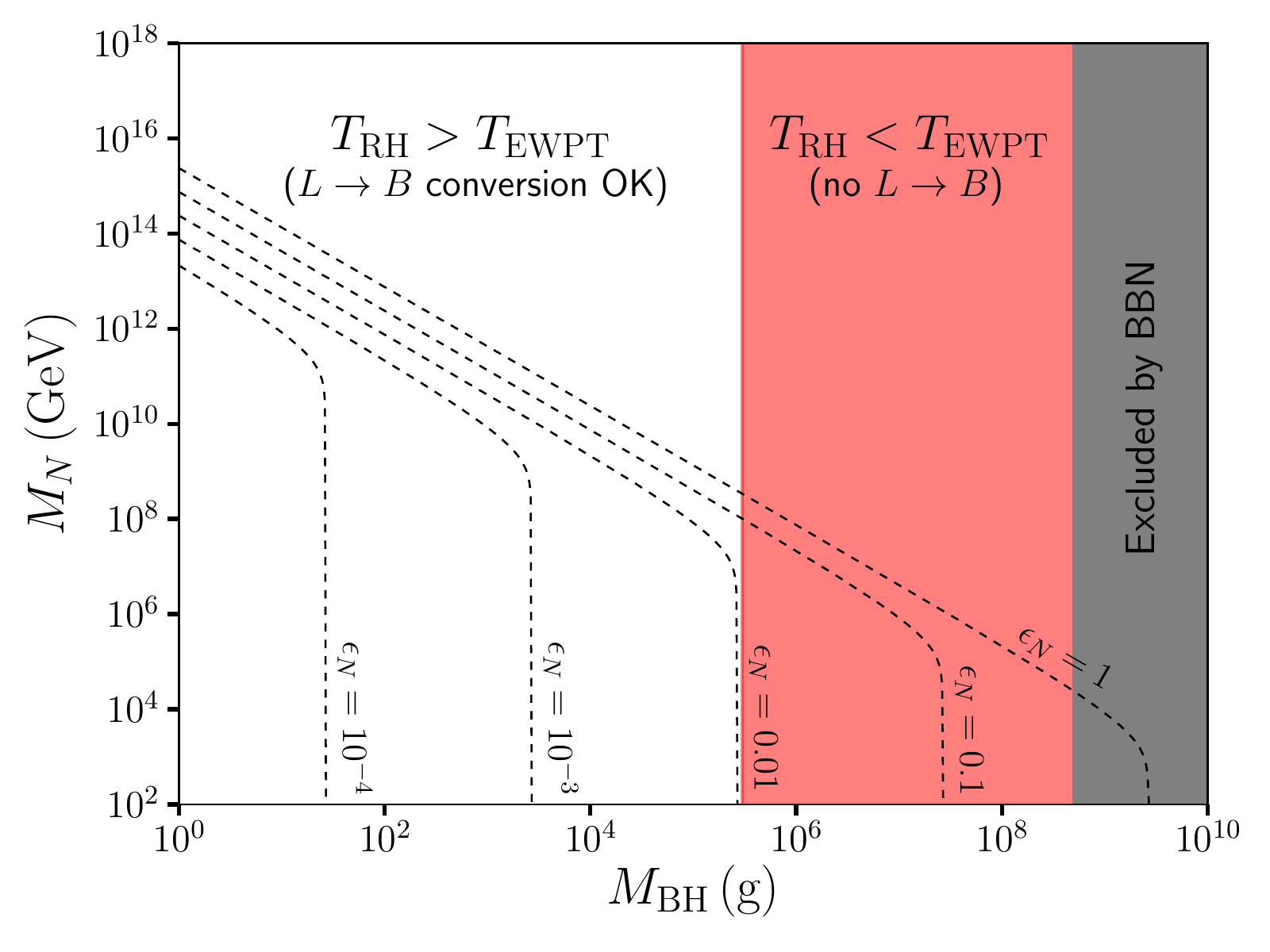} 
\caption{As in the previous figures, but for the case of leptogenesis. The dashed lines represent values of the right-handed neutrino mass and initial black hole mass that generate the observed baryon asymmetry, for various values of $\epsilon_N$. In this figure, we have assumed that black holes dominated the energy density of the universe prior to their evaporation. In the red region, the universe is reheated by black hole evaporation to a temperature below than that of the electroweak phase transition, making it impossible for sphalerons to convert the lepton asymmetry into a baryon asymmetry. Black holes lighter than $M_{{\rm BH}, i} \lsim 3 \times 10^5 \, {\rm g}$ could easily generate the observed baryon asymmetry through the Hawking radiation of lepton number violating right-handed neutrinos.}
\label{fig4}
\end{figure}

In leptogenesis, a lepton number asymmetry is generated through the decays of right-handed neutrinos, which is then converted into a baryon asymmetry through processes involving sphalerons. If black holes are present and evaporating in the early universe, they could radiate these right-handed neutrinos, generating an asymmetry in a similar fashion to that described in the previous section. There are some key differences, however, between leptogenesis and baryogenesis within this context. In particular:
\begin{itemize}
\item{Unlike baryogenesis, leptogenesis requires sphalerons to transfer the lepton asymmetry produced through right-handed neutrino decays into a baryon asymmetry. Thus, in order for the right-handed neutrinos radiated from black holes to produce the observed baryon asymmetry, the black holes must be light enough to reheat the universe to a temperature {\it above} the electroweak phase transition, $M_{\rm BH} \lsim 3 \times 10^{5} \, {\rm g}$. This is opposite of the requirement found in the case of black hole assisted GUT baryogenesis.}
\item{The right-handed neutrinos do not induce proton decay, and thus are free to be relatively light, unlike GUT Higgs and gauge bosons.}
\item{Due to its scalar nature, the contribution to $g_{\star, H}$ from the Higgs triplet is somewhat enhanced relative to that from right-handed neutrinos (by a factor of 1.82 per degree-of-freedom).}
\end{itemize}

In Fig.~\ref{fig4}, we summarize the parameter space in which black holes can generate the baryon asymmetry through the lepton-number violating decays of right-handed neutrinos. Here, we have considered three generations of right-handed neutrinos of mass, $M_N$. Black holes lighter than $M_{{\rm BH},i} \lsim 3 \times 10^5 \, {\rm g}$ could easily generate the observed baryon asymmetry in this class of scenarios.

Lastly, we will comment that whereas constrains from the CMB and proton decay have ruled out or tightly constrained many models of GUT baryogenesis (thus motivating a role for black holes, as discussed in this paper), leptogenesis models have generally not been impacted by such constraints and remain broadly viable. From this perspective, the presence of primordial black holes does not seem to be a necessary (or even helpful) feature for leptogenesis.

\section{Discussion and Summary}
\label{sec:summary}

A significant abundance of black holes may have been formed in the early universe, either as a consequence of inflation or in subsequent phase transitions. Given that the energy density of black holes evolves as matter rather than radiation, their fractional abundance will increase as the universe expands, until they evaporate through the emission of Hawking radiation. From this perspective, it is well motivated to consider scenarios in which the total energy density of the early universe was dominated by black holes prior to their evaporation. Such black holes could potentially have played an important role in establishing the current composition of the universe. In particular, it has been argued that the Hawking evaporation of primordial black holes may have produced the universe's dark matter abundance~\cite{Fujita:2014hha,Lennon:2017tqq,Allahverdi:2017sks,Morrison:2018xla,Hooper:2019gtx,Masina:2020xhk}, an observable density of dark radiation~\cite{Hooper:2019gtx,Hooper:2020evu,Masina:2020xhk}, or even a stochastic background of high-frequency gravitational waves~\cite{Hooper:2020evu,Inomata:2020lmk}. In this paper, we have considered the possibility that the baryon asymmetry of the universe may have been generated through the Hawking evaporation of black holes (see also, Refs.~\cite{Barrow:1990he,Baumann:2007yr,Morrison:2018xla}), focusing the mechanism of GUT baryogenesis.

In Grand Unified Theories (GUTs), a net baryon abundance can be generated through the out-of-equilibrium decays of gauge or Higgs bosons, whose couplings violate the conservation of CP and baryon number. In standard GUT baryogenesis, these particles are generated in the very early universe, at temperatures not far below the GUT scale. Recent constraints on the scale of inflation and the subsequent temperature of reheating, however, have begun to put pressure on this class of scenarios. In contrast, black holes could produce such particles through Hawking evaporation at significantly later times, regardless of the masses of the particles or the temperature of the early universe. Furthermore, by producing the baryon number violating particles through Hawking evaporation, the decays of these particles will automatically occur when they are out-of-equilibrium with the thermal bath, naturally satisfying Sakarkov's third condition.

In typical GUTs, the baryon-number violating Higgs color triplet is taken to be very heavy, with a mass not far below the GUT-scale. In this study, we have also considered the possibility that the Yukawa couplings of the Higgs triplet could be suppressed, allowing these particles to be relatively light without inducing an unacceptable rate of proton decay. In this class of scenarios, and for black holes with masses in the range of $M_{\rm BH} \sim 10^5-10^9 \,{\rm g}$, the baryon asymmetry could be generated after the electroweak phase transition, naturally evading the effects of sphalerons which would ordinarily wash out any net baryon-plus-lepton number that was produced at early times. We have also identified interesting regions of parameter space in which the Higgs triplets can be long-lived, and come to dominate the energy density of the universe prior to their decays.

\acknowledgments

We would like to thank Nikita Blinov, Djuna Croon, and Seyda Ipek for helpful discussions. This work has been supported by the Fermi Research Alliance, LLC under Contract No. DE-AC02-07CH11359 with the U.S. Department of Energy, Office of High Energy Physics. 

\bibliography{PBH_BAU_final2}

\appendix
\section{Long-Lived Triplet Domination}
\label{appendix}

\subsection{Conditions for Triplet Domination}
\label{A1}

Throughout this paper, we have largely focused on scenarios in which rapidly evaporating primordial black holes dominate the energy density of the early universe prior to their evaporation. Upon the completion of their evaporation, the Hawking radiation results in a radiation dominated universe (along with additional heavy GUT bosons) that subsequently evolves according to standard cosmological assumptions.  This sequence of events holds true as long as the heavy triplets decay promptly and 
contribute negligibly to the entropy of the surrounding radiation bath. In the case shown in Fig. \ref{fig2LL}, however, the Yukawa couplings of the Higgs triplet are highly suppressed, making these particles long-lived. 

Immediately after the completion of black hole evaporation,
 the fraction of the total energy density that is in the form of triplets is given by the ratio
 of Hawking evaporation factors,
$ \rho_{\cal T}(\tau_{\rm BH}) /  \rho_{\cal R}(\tau_{\rm BH})  =    g_H^{\cal T} /    g_{\star, H} $,
 leading to the following time-dependent energy density ratio:
 \be
 \frac{\rho_{\cal T}(t) }{\rho_{\cal R}(t) } = 
  \frac{\rho_{\cal T}(\tau_{\rm BH}) }{\rho_{\cal R}(\tau_{\rm BH}) } \frac{a(t)}{a(\tau_{\rm BH})}
=   \frac{g_H^{\cal T}}{g_{\star, H} }  \frac{a(t)}{a(\tau_{\rm BH})}~,
 \ee
 where $\rho_{\cal R}$ is the energy density in radiation. Since 
 $a \propto t^{1/2}$ during this era, the time
 at which triplets begin to dominate the energy density of the universe can be written as
 \be
 t_{\cal T \rm D} = \tau_{\rm BH} \left(     \frac{g_{\star, H}}{g^{\cal T}_H }        \right)^2~,
 \ee
which is only realized if the triplets survive past this time, so we demand
\be
\tau_{\cal T} = \frac{16 \pi }{ 7 y^2 M_{\cal T}}  >
 t_{\cal T \rm D} = 
 \tau_{\rm BH} \left(     \frac{g_{\star, H}}{g^{\cal T}_H }        \right)^2
 =\frac{ 10240 \pi \mbh^3}{{\cal G} g_{\star, H}\mpl^4}\left(     \frac{g_{\star, H}}{g^{\cal T}_H }        \right)^2,
\ee
 where we have taken $y_u = y_d \equiv y$ in \Eq{Twidth} for simplicity. This implies a condition 
 on the triplet mass
 \be
 \label{mass-criterion}
M_{\cal T}  < 
 \frac{16 \pi}{7 y^2} \frac{{\cal G} g_{\star, H}\mpl^4}{ 10240 \pi \mbh^3}\left(     \frac{g^{\cal T}_H }{g_{\star, H}}        \right)^2
\approx 1.2 \times 10^{7} \, {\rm GeV} \,
\bigg(  \frac{ 10^{-7}  }{y} \bigg)^2
\bigg(  \frac{10^3 \, \rm g}{\mbh} \bigg)^3  
\bigg(  \frac{g_{\star,H}}{316} \bigg)
\bigg(  \frac{g^{\cal T}_H / g_{\star,H}}{0.065} \bigg)^2.
 \ee
 Thus for sufficiently small triplet masses and/or Yukawa couplings, the triplets dominate the energy
 density before decaying. If we fix the triplet Yukawa to saturate the proton decay bound, $y \simeq M_{\cal T}/3\times 10^{11} \, {\rm GeV}$ (as in Fig. \ref{fig2}), the criterion for triplet domination becomes
 \be
 M_{\cal T} \lesssim  2.2  \times 10^{5} \, {\rm GeV} 
 \bigg(  \frac{10^3 \, \rm g}{\mbh} \bigg)  
\bigg(  \frac{g_{\star,H}}{316} \bigg)^{1/3}
\bigg(  \frac{g^{\cal T}_H / g_{\star,H}}{0.065} \bigg)^{2/3},
 \ee
which is only satisfied in regions of parameter space in which the triplets are efficiently depleted through annihilation. In contrast, in the case considered in Fig.~\ref{fig2LL}, the Yukawa couplings of the triplet are taken to be much smaller, $y \simeq 10^{-8} \times (M_{\cal T}/3\times 10^{11} \, {\rm GeV})$, significantly relaxing the criterion for triplet domination: 
\be
 M_{\cal T} \lesssim  4.8  \times 10^{10} \, {\rm GeV} 
 \bigg(  \frac{10^3 \, \rm g}{\mbh} \bigg)  
\bigg(  \frac{g_{\star,H}}{316} \bigg)^{1/3}
\bigg(  \frac{g^{\cal T}_H / g_{\star,H}}{0.065} \bigg)^{2/3}.
 \ee
 
Along each of the contours of constant $Y_B = 8.8 \times 10^{-11}$ shown in Fig.~\ref{fig2LL}, this criterion is consistently satisfied, and thus triplets dominate the energy density of the universe prior to their decays.


\subsection{Decays During Triplet Domination}
\label{moderate}

If the triplets decay after they begin to dominate the energy density of the universe, $\tau_{\cal T} > t_{{\cal T}D}$, the 
 entropy increase is given by~\cite{Kolb:1990vq}
 \be
 \label{sratio1}
 \frac{s(\tau_{\cal T})  }{s (\tau_{\rm BH}) } \simeq 1.83 \,g_\star(T_{\rm RH})^{1/4} \frac{M_{\cal T }Y_{{\cal T}} (\tau_{\rm BH})  }{  (\mpl \Gamma_{\cal T})^{1/2} },
 \ee
 where $Y_{{\cal T}} (\tau_{\rm BH}) = n_{{\cal T}}(\tau_{\rm BH})/s(\tau_{\rm BH})$ is the initial yield of triplets and, in the approximation of instantaneous black hole evaporation, the entropy in the denominator of Eq.~(\ref{sratio1}) is written as 
\be
  s(\tau_{\rm BH}) = \frac{2 \pi^2 g_\star(T_{\rm RH})}{45} T_{\rm RH}^3
 =
\frac{2 \pi^2 g_\star(T_{\rm RH})}{45}\left(  \frac{  5 \mpl^2 }{\pi^3 g_\star(T_{\rm RH}) \tau_{\rm BH}^2} \right)^{3/4}~,
 \ee
 where we have used \Eq{TRH1}. In the $T_{{\rm BH},i} \gg M_{\cal T}$ case, the number density of triplets immediately after black hole evaporation is given by
 \be
 \label{nt1}
 n_{{\cal T}}(\tau_{\rm BH})  = \frac{ \rho_{{\cal T} } (\tau_{\rm BH})     }{     M_{\cal T}  } 
= 
 \frac{g_H^{\cal T}}{g_{\star, H} }
 \frac{ \rho_{\rm BH}(\tau_{\rm BH })  }{     M_{{\cal T}}  }     
 =
 \frac{g_H^{\cal T}}{g_{\star, H} }
 \frac{   \mpl^2  }{  6\pi    M_{{\cal T}} \tau_{\rm BH }^2  }     ,
 \ee
 where we have used the fact that $\rho_{\rm BH}(\tau_{\rm    BH}) \propto H^2 = (2/3 \tau_{\rm BH})^2$, so the final entropy at the time of triplet decay now becomes
 \be
  s(\tau_{\cal T})\simeq
  \frac{1.83 \, g_\star^{1/4}(T_{\rm RH})  }{   \Gamma_{\cal T}^{1/2} }
    \frac{   \mpl^{3/2}  }{  6\pi     \tau_{\rm BH }^2  }     
 \frac{g_H^{\cal T}}{g_{\star, H} } ,
 \ee
 and the final number density of triplets can be obtained by 
 redshifting the initial density from Eq.~(\ref{nt1}) in two steps. First we redshift up to the scale factor
 at on the onset of triplet domination, $a(t_{{\cal T}\rm D}) = a(\tau_{\rm BH})    (g_{\star,H}/g^{\cal T}_H) $,
 \be
 n_{\cal T}(t_{{\cal T}\rm D}) =  n_{{\cal T}}(\tau_{\rm BH}) \left(   \frac{g_H^{\cal T}}{g_{\star, H} }   \right)^3 ,
 \ee 
 which we then redshift further to the time of triplet decay during matter domination, $t = \tau_{\cal T} $,
\be
 n_{\cal T}(\tau_{\cal T})  =  n_{\cal T}(t_{{\cal T}\rm D})  \left( \frac{     t_{{\cal T}\rm D}       }{ \tau_{\cal T}} \right)^{2}
 = 
    \frac{   \mpl^2  }{  6\pi    M_{{\cal T}} \tau_{\cal T}^2  }     ~,
\ee 
where we have used $a \propto t^{2/3}$ and Eq.~(\ref{nt1}).
Putting this all together, the resulting baryon yield, $Y_B=\epsilon_{\cal T} n_{\cal T}(\tau_{\cal T})/s(\tau_{\cal})$, is given by
\be
\label{YB-moderate-domination}
Y_B &\simeq&
\frac{4\times 10^6 \, \epsilon_{\cal T} y_{t,b}^5   }{ {\cal G}^2        g_\star(T_{\rm RH})^{1/4}  \,  g_{\star, H}  \, g_H^{\cal T}   }   
\frac{   M_{\cal T }^{3/2}  \mbh^6  }{  \mpl^{15/2}       } \\ 
&\approx & 10^{-10} \bigg(\frac{\epsilon_{\cal T}}{10^{-3}}\bigg) \bigg(\frac{y}{10^{-8} \, y_{\rm max}}\bigg) \bigg(\frac{316}{g_{\star, H}}\bigg)\bigg(\frac{106.75}{g_{\star}(T_{\rm RH})}\bigg)^{1/4} \bigg(\frac{M_{\cal T}}{10^{12} \, {\rm GeV}}\bigg)^{3/2} \bigg(\frac{M_{\rm BH}}{10^2 \, {\rm g}}\bigg)^6, \nonumber
%
%
 \ee
which is realized along the diagonal contours of constant $Y_B$ in Fig.~\ref{fig2LL}, in which triplets dominate the energy density of the early universe but the surrounding pre-decay Standard Model entropy is still relevant.

\subsection{Way-out-of-Equilibrium Decays}
\label{way-out}
 In the limit in which the triplets are very long lived and dominate the universe's energy
 density long enough to dilute away any pre-existing entropy to negligible levels, the subsequent evolution 
 of the universe is insensitive to any initial conditions (e.g.~the mass of the black holes, or the previous radiation
 temperature). In this regime, triplet decays eventually reconstitute the thermal bath through their decays and ``re-reheat" the universe to the following temperature:
 \be
 T_{\rm RH \cal T} \simeq  \frac{ 0.55 }{g_{\star}(T_{\rm RH})^{1/4}} \left( \frac{\mpl}{\tau_{\cal T}} \right)^{1/2}
 =  \frac{ 0.55 }{g_{\star}(T_{\rm RH})^{1/4}} \left( \frac{7 y^2\mpl M_{\cal T   }        }{16\pi} \right)^{1/2}~,
 \ee
so the baryon yield can be written as~\cite{Kolb:1990vq}
 \be
 Y_B = \frac{n_B(T_{\rm RH \cal T} )}{s(T_{\rm RH \cal T})} \simeq \epsilon_T \frac{T_{\rm RH \cal T}}{ M_{\cal T} }
 \simeq
 \frac{ 0.55 \epsilon_{\cal T} }{g_{\star}(T_{\rm RH})^{1/4} M_{\cal T}  } \left( \frac{7 y^2\mpl M_{\cal T   }        }{16\pi} \right)^{1/2}
 =
 \frac{ 0.2 y \epsilon_{\cal T}  }{g_{\star}(T_{\rm RH})^{1/4}  } \left( \frac{ \mpl         }{  M_{\cal T   }  } \right)^{1/2},
 \ee
 which is independent of $M_{\rm BH}$. For a reference value of $y=10^{-8} \, y_{\rm max}$, the baryon yield becomes
 \begin{equation}
 Y_B \approx 10^{-10} \, \bigg(\frac{y}{10^{-8} \, y_{\rm max}}\bigg) \bigg(\frac{106.75}{g_{\star}(T_{\rm RH})}\bigg)^{1/4} \bigg(\frac{2\times 10^6 \, {\rm GeV}}{M_{\cal T}}\bigg)^{1/2}.  
 \end{equation}
 The above expression corresponds to the horizontal contours in Fig.~\ref{fig2LL}, along which the triplets dominate the energy density of the early universe and redshift away the preexisting Standard Model entropy to negligible levels.

\end{document}